# Laser studies of metallic artworks

E. Drakaki[1], M. Kandyla[1], E. Chatzitheodoridis[1], I. Zergioti[1], A.A. Serafetinides[1], A. Terlixi[2], E. Kouloumpi[2], A. Moutsatsou[2], M. Doulgerides[2], V. Kantarelou[3], A. Karydas[3], C. Vlachou-Mogire[4]

[1]Physics Department, National Technical University of Athens, 15780 Athens, Greece,
[2]National Gallery of Alexandros Soutzos Museum, Conservation Department, Laboratory of Physicochemical Research, 1 I. Michalacopoulou Str., 11601 Athens, Greece,
[3]Institute of Nuclear Physics, Laboratory for Material Analysis, NCSR "Demokritos", GR- 15310 Ag. Paraskevi, Greece,
[4]Numismatic Museum of Athens, 12 Eleftheriou Venizelou Avenue, 10671, Athens, Greece.

**Abstract**

Museum curators and archaeologists use analytical science to provide important information on artworks and objects. For example, scientific techniques provide information on artwork elemental composition, origin and authenticity, and corrosion products, while also finding use in the day-to-day conservation of many historical objects in museums and archaeological sites around the world. In this work two special cases are being discussed.

In the first part of our work, physicochemical studies of an icon on a metal substrate were carried out using non-destructive, qualitative analysis of pigments and organic-based binding media, employing various microscopic and analytical techniques, such as Optical Fluorescence Microscopy, XRF, and Gas Chromatography. In the second part of our work, laser cleaning of late Roman coins has been performed using a Q-switched Nd:YAG laser (1064 nm, 6 ns) and a GaAlAs diode laser (780 nm, 90 ps). The corrosion products have been removed, while we observe increased concentrations in Ag, which is the main material of the silvering plating found in late Roman coins.

## 1. Introduction

Laser technology has great potential for the development of effective conservation procedures due to its controllability and reproducibility. Lasers can remove unwanted layers that conventional techniques cannot remove safely. However, laser cleaning is an intrusive technique, which should be complemented by detailed control methods. Museum curators and archaeologists use analytical scientific techniques on parts of artworks (*e.g.* on coatings, paints and pigments), in order to determine the place of origin, authenticity, corrosion products and their cause, and for day-to-day conservation of priceless historical objects in museums and archaeological sites around the world. In this work two special cases of metallic artworks are being studied.

Metals have never been used extensively as painting supports. In ancient times they were probably highly valued for other purposes and were not available in large sheets or panels [1]. Copper plates have been used as supports for oil and oil resin paintings in the middle ages, particularly in Holland [2]. However, Kommanecky *et al*. [3] demonstrate that this painting technique has its origins in sixteenth-century Italy. Vasari [4] reported that Sebastiano del Piombo made paintings on silver, lead, and copper, presumably around 1530, although no works on copper survive. Around this time, Antonio Correggio painted a Penitent Magdalen on copper and it seems that Parmigianino also did a painting on copper [3]. This practice was soon adopted by Northern European artists. The practice of painting on copper was widespread in Europe during the seventeenth century and Antwerp played a major role in this. The popularity of painting on copper would gradually wane after 1650 [3]. Paintings in general and especially paintings on metal substrates, present a complicated structure, which originates from i) the multilayer painting technique and ii) the heterogeneity of the materials constituting the successive layers. This complexity requires the identification of materials and painting technique used for an artwork before proceeding with conservation and restoration treatments. Regarding laser cleaning of paintings on metal, earlier application of a diagnostic procedure is considered indispensable in order to specify the relevant physicochemical parameters and define the aim of the treatment (removal of metal-oxide encrustations, removal of superficial varnish layer, etc). In the first part of this research, physicochemical studies of an icon on a metal substrate were carried out. For this purpose, several high performance analytical techniques were employed.

Another set of metallic artworks is the case of late Roman coins, the surface of which is covered by a thin silver plating layer. Political problems in the Late Roman Empire caused significant changes in the coin technology. The silver content dropped severely and a new technology was introduced, which was applied to all the mints operating around the Empire. For the production of these coins, copper based quaternary alloys were used and their surface was covered by a few microns thick, amalgam silvering plating [5]. Due to the small thickness of the silvering plating, the conservation treatment of such coins has been challenging.

Both mechanical and chemical cleaning result in the damage or complete destruction of the thin silver layer. The use of laser technology for cleaning works of art has a wide range of applications, including metallic objects [6-12]. Besides the established use of lasers emitting pulses of nanosecond duration for cleaning applications [6-12], picosecond lasers present an interesting new possibility for demanding conservation cases as well [13]. Picosecond laser pulses are important in cases where the unwanted surface layers are very thin, and thus it is necessary to ensure that the mean removed depth per laser pulse, as well as the thermal load to the substrate, is the minimum possible.

In the second part of this work we investigated the influence of pulse duration on laser cleaning of thin silver plating layers found in late Roman coins. Cleaning tests were performed using a Nd:YAG (1064 nm - 6 ns) and a GaAlAs diode (780 nm - 90 ps) laser system. The cleaning results on the plated areas were examined by high resolution optical microscopy and SEM-EDX analysis. Comparative results of this study are presented here.

**2. Materials and Methods**

The painting used in this work constitutes of a metal plate support on which painting layers are applied (Fig.1). A thick layer of glossy varnish is observed on the surface. The backside of the metal plate is also covered with a light purple colored layer. The depicted subject is "Madonna with a Child" in a three-quarter style depiction.

Several Roman coins from Cope's Archive were also used for the purposes of this research. Although experiments were performed on several coins that were issued during the period of Licinius, in 315 AD from the mint of Alexandria, here we will concentrate on one representative coin, labelled NMW56. The coin NMW56 was issued from the Gallic Empire in 260-269 AD at the mint of Cologne. This was a period of change so the silver content is higher than the other coins we have used. Although the silver content is high, a silver plating layer is also present on the surface. The coin diameter is 20 mm and the composition of the substrate alloy, according to Cope's analytical results, is shown in Table 1. (The code NMW56 belongs to Cope's coin reference system, which is followed by most researchers).

Optical microscopy was applied in both cases. All samples selected from the painting for examination with Visible Light Microscopy (VLM) and Fluorescent Light Microscopy (FLM), were encubed in a polyesteric resin and polished with various grain size sand papers [14]. A Leica DM/LM microscope, equipped with a set-in system of reflected fluorescence, and a Leica DC 300 F camera were used for examination of the encubed samples. For the observation of UV fluorescence, a filter cube - type A (emission: 425 nm, excitation: 340-380 nm) by Leica was used. Optical microscopy examination of the coin revealed that the surface was covered with a corrosion layer. Although there is a large amount of available information on the corrosion of binary and ternary copper alloys [15-17], the composition and structure of corrosion layers of quaternary copper alloys have only been investigated lately [19-22]. Metallographic and microscopic examination revealed that the surface of the coin was covered mainly with copper corrosion products [23]. The plating was made of a silver layer with a thickness of 1 μm, which had suffered from corrosion in some areas [23].

The painting was further analyzed by a model 8700 Perkin Elmer gas chromatograph with a flame ionization detector [24]. Helium (99.999%) was used as the carrier gas. The chromatographic separations were achieved on a 15 m long column from Restek, with an internal diameter of 0.25 mm (RTX-1701), using the following temperature protocol: the initial temperature was set at 70 $^{o}$C for 1 min, and then it was increased by 27 $^{o}$C/min up to 250 $^{o}$C, where it was maintained for 10 min. The injector and detector temperatures were 240 $^{o}$C and 260 $^{o}$C, respectively. The helium head pressure was 17 psig. The split ratio was 20:1. The relative standard deviation (RSD) was around 5% on 5 consecutive runs for each sample.

For surface chemical analysis of the painting, a portable μ-XRF spectrometer, developed in N.C.S.R. "Demokritos", with a Rh-anode air cooled X-ray tube, a silicon drift detector (Bruker-AXS), an optical element (polycapillary lens, IFG), a laser pointer, and a colour CCD camera was used. In addition, three independent stepping motors, coupled to the spectrometer head, allowed for three-dimensional movement, setting precisely the analysis spot at the focal length of the polycapillary lens and providing the capability of advanced elemental mapping studies.

Two laser systems were employed for surface cleaning of the NMW56 coin. The first one is a Q-switched 1064 nm Nd:YAG laser with a pulse duration of 6 ns, a repetition rate of 1 Hz, and an output pulse energy of 63 mJ. For this system the beam is focused using an 85 mm focal length cylindrical lens. The focal point is behind the sample surface and the beam area at the surface of the sample is 6.3 mm$^2$, yielding an incident fluence of 1 J/cm$^2$. The second system is a GaAlAs external cavity diode laser, with a pulse duration of 90 ps, a repetition rate of 490 MHz, central wavelength at 780 nm, and output pulse energy at 0.25 nJ. Due to the low pulse energy, for this system the beam is focused very tightly onto the sample surface by a NA 0.55 microscope objective lens, yielding an incident fluence of 0.06 J/cm$^2$. In both sets of experiments the samples were mounted on x-y-z micro-adjustable stages.

Both the painting and the coin were characterized by scanning electron microscopy – energy dispersive analysis (SEM-EDX). A scanning electron microscope (FEI Quanta Inspect operated at 25 kV) with an EDAX Genesis ultra-thin film window energy dispersive X-ray micro-analyser was used.

## 3. Results and discussion

**A. Metallic painting**

Seven cross sections were taken for examination with VLM and FLM. Six cross sections were taken from the front side of the icon and one from the back side, where the metal plate is covered entirely by a light purple layer. The images for two cross sections from the front side of the icon are shown in Figs. 2 and 3.

From microscopic examination of the six samples taken from the front side of the icon we conclude that the layer structure of the paint surface consists of: 1) a homogenous brown shaded organic substance (right above the metal plate), 2) an orange-brownish shaded ground layer which includes white semitransparent grains, brown small sized grains, orange very small sized grains, and a few black grains, 3) the paint layers (one or more depending on the location of the cross section), and 4) and 5) one or two varnish layers.

At the back side of the icon we observe a similar layer structure, which consists of: 1) a homogenous organic substance (right above the metal leaf), 2) an orange-brownish shaded ground layer which includes white semitransparent grains, brownish shaded transparent rectangular grains, very small spherical orange grains, a few black grains, small brown spherical grains, and very few green grains, 3) a light purple shaded layer, and 4) a layer of varnish.

Below we can see gas chromatography data for two icon samples, where each peak corresponds to a specific substance (Figs. 4 and 5). Table 2 shows fatty acid ratios, as determined by the data shown in Fig. 4, along with fatty acid ratios of known substances. These ratios are used for characterisation of the organic binder (the material that binds the pigment particles together and to the substrate). In this case, the results indicate the presence of linseed oil [25]. Table 3 shows fatty acid ratios, as determined by the data shown in Fig. 5, along with fatty acid ratios of known substances. Again, the results indicate the presence of linseed oil as a binder.

Based on XRF results (not shown here), we conclude that the metallic painting support is made of pure zinc, since no other typical zinc alloy metal (Al, Cu, etc.) is detected. Moreover, Pb, present at the relative XRF spectrum, is attributed to lead white used for the background pigment mixture. Although zinc plates are not among the common metals traditionally used as supports for oil paintings [1, 2, 3], they have been extensively used in recent years for decorative purposes due to their excellent resistance to atmospheric changes [25].

However, the presence of Pb, even in small quantities, leads to the corrosion of zinc grains in the presence of humidity [26]. EDX analysis identified elements that indicate the use of the following pigments in the paint layers: Cinnabar (HgS), Viridian ($Cr_2O_3$), Red Ochre ($Fe_2O_3$), Raw Umber ($Fe_2O_3$+$MnO_2$+$H_2O$, clay, etc.), and White lead (PbSO4·PbO). Further research is still in progress about the probable use of orpiment or realgar, the identification of the blue pigment (probably organic), etc.

**B. Roman coin**

Optical microscope images revealed the inhomogeneity of the corrosion layer and the varying thickness of the encrustation between the peripheral area of the coin, the centre, and the letter area, which made the controlled removal of the unwanted corrosion layer more challenging. The microscope images illustrate a defect-rich, inhomogeneous and porous corrosion layer, which forms over the surface of the silver plating in a looser manner, as reported in the literature [27]. Time-induced deterioration effects can be seen on the silver plating, in areas where fragments of the silver are evident. Below we present the cleaning results for each laser system separately.

**a. Nd:YAG (1064 nm - 6 ns) laser cleaning**

After several trials on the coin surface using different laser fluences, the incident fluence of 1 J/cm$^2$ was chosen as optimum. Laser fluences below this value didn't remove the corrosion products, while laser fluences above this value were causing damage on the coin substrate. The fluence of 1 J/cm$^2$ was used in order to clean a larger area of the coin by scanning the laser beam across 10 points, each 500 μm away from the other. This procedure created an overlap of about 50% between adjacent laser spots. Only one laser pulse was incident on each spot. Figure 6a shows a SEM image of the cleaned part of the coin surface, appearing brighter, surrounded by the darker, non-cleaned encrustation. Figure 6b shows the cleaned part of the coin surface in higher magnification. As we can see in Fig. 6b, the cleaned part of the coin surface is slightly inhomogeneous, covered by spherical beads and longer, hair-like structures. EDX measurements reveal that the spherical formations consist of Cu, which melts and resolidifies in beads, while the longer structures consist mainly of Ag, which emerges as the corrosion layer is removed by the laser beam.

Quantitative EDX measurements were performed on the non-cleaned part of the coin surface, in order to identify the corrosion products. The most dominant element is Cu, indicating that the Ag layer is covered by Cu corrosions. A strong presence of O indicates the surface is covered by Cu oxides. Other corrosion elements, found on the coin surface, include Cl, Na, S, Si, K, Ca, Pb, and Mg. EDX measurements were also performed on the clean part of the coin surface, on the points A and B, and on the areas C, D, and E, as indicated in Fig. 6b. The results are presented in Table 4, where the weight percentage of each element found on the clean coin surface is reported. Table 4 shows that all the corrosion products found on the coin surface disappear upon irradiation with the laser beam. At the same time the Ag layer, not originally present on the surface, is revealed. In certain areas, as in area D, the Ag concentration is 90%, which means almost all the corrosion is removed.

**b. Diode GaAlAs (780 nm - 90 ps) laser cleaning**

A microscope objective was used in order to focus the output of the diode laser on the sample surface. This was the only way to achieve cleaning results due to the low pulse energy of this laser. The laser beam was scanned across the sample surface, producing grooves of clean areas of 10 μm width each, as shown in the SEM image in Fig. 7a. Approximately $10^8$ pulses are incident on each spot due to the particularly high repetition rate of this laser system (490 MHz). Figure 7b shows a higher magnification SEM image of one of the grooves shown in Fig. 7a, while quantitative results following EDX measurements at points A and B are presented in Table 5.

The results presented in Table 5 reveal that partial cleaning of the sample surface has been achieved in the case shown in Fig. 7b. None of the corrosion elements found initially on the coin surface appears in the quantitative results presented in Table 5, apart from a minor Si quantity. Furthermore, the Ag layer has been revealed, although there are still significant amounts of Cu present. Not all fluences of the diode laser were capable of cleaning the coin surface. We observed that the threshold for coin cleaning was at 0.06 J/cm$^2$ for $10^8$ incident laser pulses, where removal of corrosion elements is possible and visible changes take place. The number of pulses is important because, for this laser system, the time interval between successive pulses is approximately 2 ns, which is much shorter than the time required for heat to diffuse out of the focal volume. Therefore, energy accumulates in and around the focal volume, causing the removal of corrosion elements, even for this low pulse fluence. For the emergence of the Ag layer, higher fluences or a bigger number of pulses are required.

## 4. Conclusions

In the case of the Roman coins, the corrosion products, including mainly Cl, Na, Si, S, K, Ca, and Mg, have been successfully removed. Specifically, although all of these elements appear on various test sites on the non-cleaned parts of the coins surface, they almost completely disappear on the laser-processed parts of the coins surface. Furthermore, on the cleaned areas of the coins, we observe increased concentrations in Ag, which is the material that was originally used by the Roman mints as the surface layer.

The Nd:YAG laser system produced better results because not only it removed the corrosion elements from the coins surface, it also revealed successfully the Ag layer buried underneath. The Nd:YAG laser system produces enough energy in order to optimize the cleaning process. The picosecond diode laser system produces significantly smaller energy per pulse. However, even for the low fluences of the diode laser, the corrosion elements were successfully removed from the sample surface due to the large number of pulses. This shows that the corrosion products are not strongly bonded to the surface of the coin. The silver plating remained buried underneath layers of copper corrosion, given there was not enough fluence to ablate copper corrosions away. In conclusion, the optimum laser fluence for cleaning late Roman coins lies in the vicinity of 1 J/cm$^2$, while for fluences as low as 0.06 J/cm$^2$ and a high repetition rate the corrosion elements disappear, however the silver plating remains buried underneath layers of copper, which require higher fluences in order to be removed.

Regarding the icon on metal substrate, the aim of this work was to identify the materials and painting technique used by the artist, in order to specify the physicochemical parameters involved in laser cleaning of the metal oxide encrustations and/or the superficial varnish layers.

The XRF results show that the painting support is pure zinc plate. From VLM and FLM examination of the six samples originating from the front side of the icon, we conclude the general layer structure consists of: a) an homogenous organic substance (right above the metal plate), b) a brownish coloured ground layer, c) the paint layers, and d) one or two varnish layers. At the back side of the metal plate a similar layer structure is observed: a) an homogenous organic substance (right above the metal leaf), b) a brownish shaded ground layer, c) a light purpled shaded layer, and d) a layer of varnish. EDX and XRF analysis identified elements that indicate the use of the following pigments in the paint layers: Cinnabar (HgS), Viridian ($Cr_2O_3$), Red Ochre ($Fe_2O_3$), Raw Umber ($Fe_2O_3+MnO_2+H_2O$, clay, etc.), and White lead ($PbSO_4 \cdot PbO$). Further research is still in progress about the probable use of orpiment or realgar, the identification of the blue pigment (probably organic), etc. The ratios of the fatty acids present in the gas chromatograms indicate the use of linseed oil.

**Acknowledgements**: This research is financially supported by John S. Latsis Public Benefit Foundation, Research Projects 2009 (project entitled "Labelling ancient coins from collections of the Athens Numismatic Museum and cleaning them with the use of laser beams"). Finally we would like to thank the Department of Coins and Medals, British Museum, for allowing access to Cope's Archive.

**References**

1. R.J. Gettens, L.S. George: Painting Materials, A short encyclopedia (Dover Publications NY 1966)
2. R. Mayer: The Artist's Hanbook of Materials and Techniques (Faber and Faber London 1991)
3. A.K. Komanecky, I. Horovitz, N. Eastaugh: Antwerp artists and the practice of painting on copper, in Preprints of the IIC Dublin Congress: Painting techniques, history, materials and studio practice (IIC, Reedprint ltd., Berkshire 1998)
4. G. Vasari: The Lives of the Painters; Sculptors & Architects (BiblioLife 2008)
5. C. Vlachou, G. McDonnell. R. Janaway,, Conservation Science, 236 (2003)
6. E. Drakaki, A.G. Karydas, B. Klinkenberg, M. Kokkoris, A.A. Serafetinides, E. Stavrou, R. Vlastou, Ch. Zarkadas, Appl. Phys. A **79**, 1111 (2004)
7. C. Vlachou-Mogire, E. Drakaki, A.A. Serafetinides, I. Zergioti, N. Boukos, Proc. of SPIE **6604**, 66040W (2007)
8. R. Pini, S. Siano, R. Salimbeni, M. Pasquinucci, M.Miccio, J. Cult. Heritage **1,** S129 (2000)
9. S. Siano, R. Salimbeni, Stud.Conserv. **46**, 269 (2001)
10. S. Siano, R. Salimbeni, R. Pini, A. Giusti, M.Matteini, J. Cult. Heritage **4**, S140 (2003)
11. S. Siano, F. Grazzi, Il Nuovo Cimento C **30,** 123 (2007)
12. S. Siano, F. Grazzi, V.A.Parfenov, J. Opt. Techn. **75,** 419 (2008)
13. V.V. Golovleva, M.J. Gresalfia, J.C. Millera, G. Romerb, P. Messierc, J. Cult. Heritage **1**, S139 (2000)
14. A. Alexopoulou-Agoranou, Y. Chryssoulakis, Exact Sciences and Works of Art (in Greek), Gkoni Publications, Athens, Greece(1993)
15A. Dorigo, C. Fiaud, J.-P. Labbé, L. Robbiola, P. Brunella, H. Böcking: Characterization of the corrosion structures of Roman copper alloys by SEM and EDSX IMACO: Improvement of measurements on Archaeological Copper Alloys for Characterization and Conservation (1998)
16. I.D. MacLeod, S. Pennec: ICOM Committee for Conservation, 9th Triennial Meetings, Dresden, ICOM, **2**, 732 (1990)

17. L. Robbiola: Caractérisation de l'altération de bronzes archangéliques enfouis a partir d'un corpus d'objects de l'age du bronze. Mécanismes de corrosion, PhD Thesis (Paris VI, 1990)
18. L. Robbiola, I. Queixalos, L-P. Hurtel, M. Pernot, C. Volfovsky, Stud. Conserv. **33**, 205 (1988)
19. C. Chiavari, A. Colledan, A. Frignani, G. Brunoro, Mat. Chem. Phys. **95,** 252 (2006)
20. L. Morselli, E. Bernardi, C. Chiavari, G.Brunoro**,** Appl. Phys. A **79**, 363 (2004)
21. G.M. Ingo, E. Angelini, T. De Caro, G. Bultrini, Appl. Phys. A **79**, 171 (2004)
22. A.G. Nord, E. Mattsson and K. Tronner, Prot. Met. **4**, 309 (2005)
23. C. Vlachou: The manufacturing and plating technology used in the production of mid-3rd/4th century AD Roman coins - An analytical study, PhD thesis (Department of Archaeological Sciences, University of Bradford, 2004)
24. D. Rood: The Troubleshooting And Maintenance Guide For Gas Chromatographers (Wiley-VCH Verlag, GMBH & Co, KGaA Weinheim Germany 2007)
25. E. Kouloumpi, G. Lawson, V. Pavlidis, Anal Bioanal Chem **387**, 803 (2006)
26. Y.Chryssoulakis, D. Pantelis, Science and technology of metallic materials (in Greek), Papasotiriou Publications, Athens, Greece (1996).
27. E.A.O. Saettone, J.A.S da Mattal, W. Alva, J.F.O. Chubaci, M.C.A. Fantini, R.M.O. Galvao, P. Kiyohara, M.H.Tabacniks J. Phys. D: Appl. Phys. **36**, 842 (2003)

# Figure captions

**Figure 1:** "Madonna with Child", painting on metal substrate.

**Figure 2:** Visible light microscopy image of a cross section from the green area at the bottom end of Madonna's cloak, magnification 500X.

**Figure 3:** Visible light microscopy image of a cross section from the area of the right hand of Madonna, magnification 500X.

**Figure 4:** Gas Chromatograph of a sample taken from the icon shown in Fig.1.

**Figure 5:** Gas Chromatograph of a sample taken from the icon shown in Fig.1.

**Figure 6:** SEM images of the coin NMW56 surface, irradiated by an Nd:YAG laser at 1064 nm, pulse duration 6 ns, incident fluence 1 $J/cm^2$. a) Overview of the cleaned part of the coin surface, b) Higher magnification image of the clean area of the coin. EDX analysis was performed at points A and B, indicated by arrows, and areas C, D, and E, indicated by squares. The EDX results are presented in Table 4.

**Figure 7:** SEM images of the coin NMW56 surface, irradiated by a diode GaAlAs laser at 780 nm, pulse duration 90 ps, incident fluence 0.06 $J/cm^2$. a) Overview of the cleaned part of the coin surface, b) Higher magnification image of the clean area of the coin. EDX analysis was performed at points A and B, indicated by the arrows. The EDX results are presented in Table 5.

**Table captions**

**Table 1.** The alloy content of coin NMW56 according to Cope's analytical results (Data from Cope's Archive).

**Table 2.** Relative ratios of the fatty acids of four reference samples and sample 1, which is taken from the icon surface. The gas chromatograph of sample 1 is shown in Fig. 4.
**Artificial ageing combining exposure to temperature, UV and four years of natural ageing.

**Table 3.** Relative ratios of the fatty acids of four reference samples and sample 2, which is taken from the icon surface. The gas chromatograph of sample 2 is shown in Fig. 5.
**Artificial ageing combining exposure to temperature, UV and four years of natural ageing.

**Table 4**. Quantitative analysis of the percentage elemental concentration of the points and areas indicated in Fig. 6b.

**Table 5**. Quantitative analysis of the percentage elemental concentration of the points indicated in Fig. 7b.

# Tables

### Table 1

| Coin | Ag | Cu | Sn | Pb |
|---|---|---|---|---|
| NMW56 | 16.56 | 83.44 | - | - |

### Table 2

| Sample | C9/C16 | C18:1/C18 | C16/C18 |
|---|---|---|---|
| Egg yolk** | 0.07 ± 0.02 | 0.2 ± 0.2 | 2.4 ± 0.1 |
| Linseed oil** | 1.00 ± 0.05 | 1.1 ± 0.5 | 1.6 ± 0.1 |
| Egg yolk-Linseed oil** | 0.5 ± 0.2 | 0.20 ±0.05 | 1.8 ± 0.4 |
| Stand oil** | 0.3 ± 0.1 | 2.1 ± 0.1 | 1.2 ± 0.1 |
| Sample 1 | 2.38 | 0.18 | 1.26 |

### Table 3

| Sample | C9/C16 | C18:1/C18 | C16/C18 |
|---|---|---|---|
| Egg yolk** | 0.07 ± 0.02 | 0.2 ± 0.2 | 2.4 ± 0.1 |
| Linseed oil** | 1.00 ± 0.05 | 1.1 ± 0.5 | 1.6 ± 0.1 |
| Egg yolk-Linseed oil** | 0.5 ± 0.2 | 0.20 ±0.05 | 1.8 ± 0.4 |
| Stand oil** | 0.3 ± 0.1 | 2.1 ± 0.1 | 1.2 ± 0.1 |
| Sample 2 | 0.5 | 0.2 | 1.3 |

### Table 4

| Area | Ag | Cu |
|---|---|---|
| A | 47.36 | 52.64 |
| B | 45.59 | 54.41 |
| C | 14.90 | 85.10 |
| D | 90.47 | 9.53 |
| E | 17.07 | 82.93 |

### Table 5

| Area | Ag | Cu | Si |
|---|---|---|---|
| A | 24.61 | 75.39 | 0 |
| B | 2.08 | 96.99 | 0.92 |

# Figures

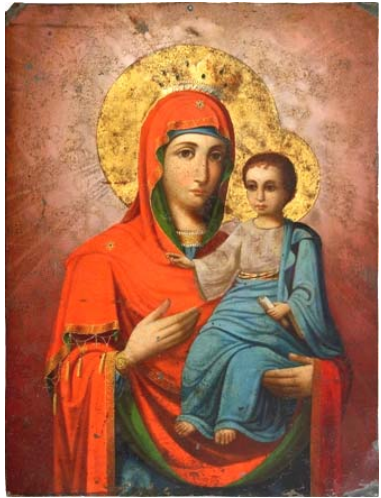

**Figure 1**

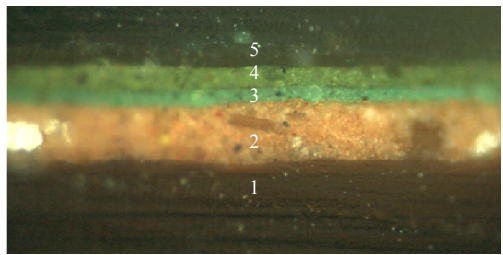

**Figure 2**

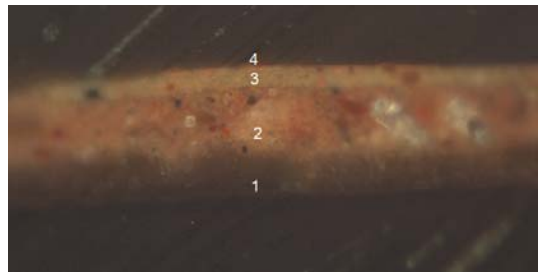

**Figure 3**

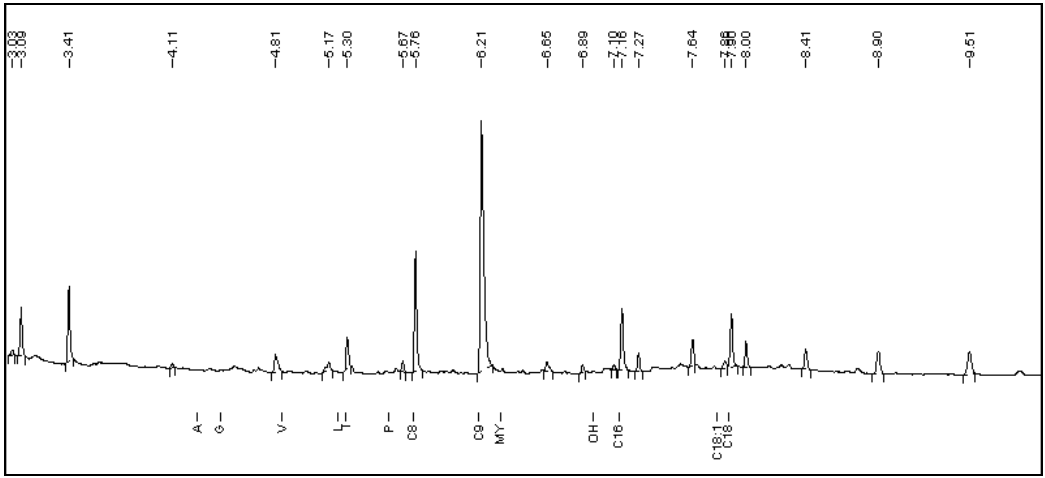

**Figure 4**

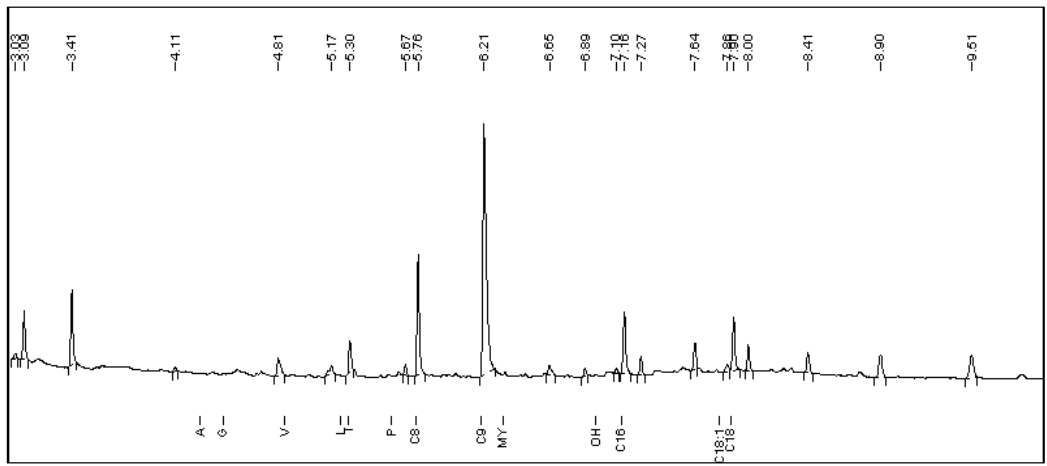

**Figure 5**

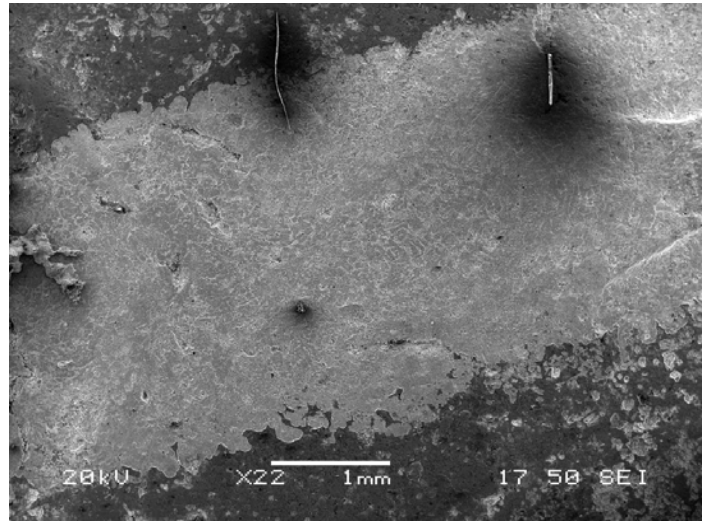

**Figure 6a**

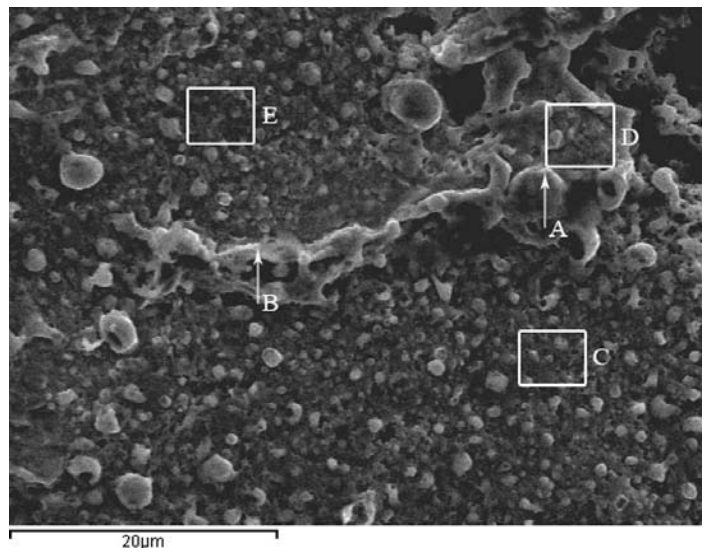

**Figure 6b**

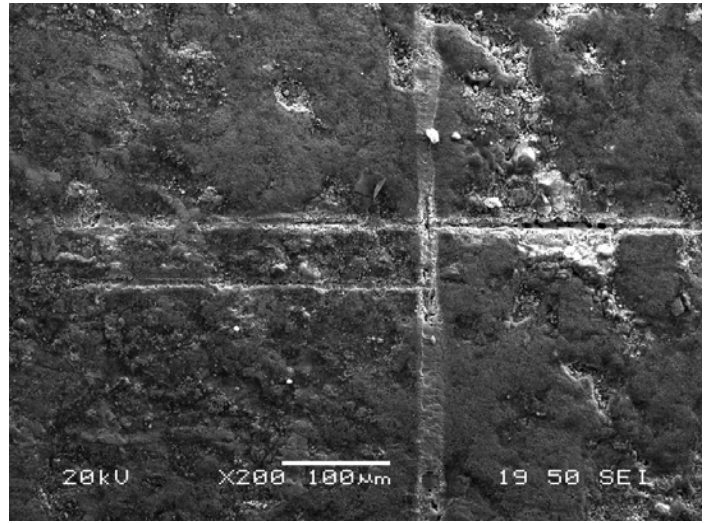

**Figure 7a**

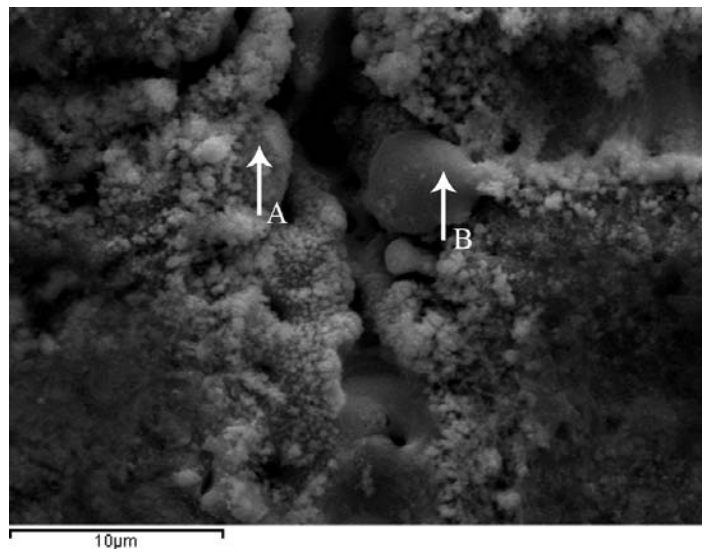

**Figure 7b**